\documentstyle[12pt]{article}

\def\tdfullfigure #1 #2 #3 #4 #5 #6
  {\begin{figure}\vspace*{#3 cm}\vspace{12.3cm}
   \tdpsinput #1 #6 #5 #3
   \vspace{-4.0cm}\vspace{#4 cm}
   \caption[]{#2}\end{figure}
   \typeout{TDINPUT: FIGURE \thefigure. produced with file #1}}
\def\tdf #1 #2 #3 #4 #5 #6
  {\begin{figure}\vspace*{#3 cm}\vspace{12.3cm}
   \tdpsinput #1 #6 #5 #3
   \vspace{#4 cm}
   \caption[]{#2}\end{figure}
   \typeout{TDINPUT: FIGURE \thefigure. produced with file #1}}
\def\tdpsinput#1 #2 #3 #4 {
   \special{ps::
-1 1 scale
-90 rotate
-1700 -2342 translate
#2 #2 scale
#4 -118 mul #3 -118 mul translate
3.5 3.5 scale
      }
   \special{ps: plotfile #1 asis}
   \special{ps:: endexecute
      }}
\begin{document}

\begin{titlepage}

\vspace*{1cm}

\begin{center}
{\bf \Large Quantum Diffusion and Tunneling with
Parametric Banded Random Matrix Hamiltonians
\footnote{ To be published in a special issue on {\it Chaos and Quantum
Transport in Mesoscopic Cosmos} of the journal {\em Chaos, Solitons \&
Fractals}, July (tentatively) 1997, guest editor K. Nakamura. } } 

\vspace{1cm}

 { Aurel BULGAC$^{a,b,}$\footnote{Internet: bulgac@phys.washington.edu},
Giu DO DANG$^{b,}$\footnote{Internet: dodang@psisun.u-psud.fr} and
Dimitri KUSNEZOV$^{c,}$\footnote{Internet: dimitri@nst.physics.yale.edu.}}\\

\vspace{5mm}

{\sl $^a$ Department of Physics, FM--15, University of
Washington, Seattle, WA 98195, USA\footnote{permanent address}}\\
\vspace{3mm}

{\sl $^b$ Laboratoire de Physique Th\'eorique et Hautes Energies,\\
      Universit\'e de Paris--Sud,  B\^at. 211, 91405 Orsay, FRANCE}\\

\vspace{3mm}

{\sl $^c$ Center for Theoretical Physics,
Sloane~Physics~Laboratory,\\ Yale University, New Haven, CT 06511-8167,USA}\\

\end{center}

\vskip 2.5 cm

\begin{abstract}

The microscopic origin of dissipation of a driven  quantum many body
system is addressed in the framework of a parametric banded random
matrix approach. We find noticeable violations of the
fluctuation--dissipation theorem and we observe also that the energy
diffusion has a markedly non--Gaussian character. Within the
Feynman--Vernon path integral formalism and in the Markovian limit, we
further consider the time evolution of a slow subsystem coupled to  such
a ``bath'' of intrinsic degrees of freedom. We show that dissipation
leads to qualitative modifications of the time evolution of the density
matrix of the slow subsystem. In either the spatial, momentum or energy
representation the density distribution acquires very long tails and
tunneling is greatly enhanced.
\end{abstract}
\vspace*{\fill}\pagebreak
\end{titlepage}

\vspace{1.5cm}

\newpage

\section{ Introduction }

It is by now an accepted fact that spectral fluctuations of many quantum
systems are identical to those resulting from random matrix theory.
Random Hamiltonians have initially been introduced by Wigner to describe 
nuclear properties at excitation energies of the order of the nucleon
binding energy \cite{random}. They have found applications in many areas 
of physics. Just to cite several directions, they have been used to
describe properties of simple quantum chaotic systems \cite{bohigas},
spectra of complex molecules \cite{molecules}, condensed matter physics
\cite{mesoscopic}, nuclear properties and spectral fluctuations obtained
in QCD calculations \cite{jack}. There is enough experimental evidence
from all these physical systems to justify the assignment of the
qualifier universal to their spectral properties.

In this article  we address the problem of the evolution of a quantum
mechanical system for which the spectral fluctuations are assumed to be
universal, the average density of states has the proper thermodynamic
behaviour and which, in addition, is capable of changing its ``shape''.
By ``shape'' we understand either some parameters describing the actual
geometrical shape of the system or some other similar global
characteristics, or externally applied fields. The shape can either be
controlled by the experimenter, or can have its own quantum dynamics.
Some classic examples of the first type of system are those studied in
thermodynamics, while examples of the second type are atomic nuclei and
complex molecules. We are interested in what happens when the shape
changes at a finite rate. In other words, we are interested in the
microscopic nature and character of dissipation in finite quantum
many--body systems. 

The paper is organized as follows. In Section II we describe the
structure of the Hamiltonian and the evolution equations for driven
systems and what we often refer to as a simple quantum system in
interaction with a complex environment. The theoretical framework in
this last case was developed many years ago by Feynman and Vernon
\cite{feynman} and used extensively, particularly in condensed matter
physics, in Caldeira--Leggett type of treatments of dissipative
tunneling \cite{caldeira}. There also exists in the literature another
type of approach, somewhat more phenomenological in nature, namely the
quantum Langevin equation \cite{ford}, which is claimed to lead to a
dissipative dynamics similar to that of the more popular
Caldeira--Leggett formalism. Our main contribution consists in the
introduction of a new type of environment or ``bath'', based on a
parametric banded random matrix approach \cite{bdk_ann,bdk_mb8,bdk_pre}.
In our opinion, such an approach, besides being extremely flexible, has
a better microscopic foundation and is better suited to describe finite
many--body quantum systems. 

In Section III we study the temporal evolution of driven complex quantum
systems which we describe using a  parametric banded random matrix
approach. The shape is an externally modulated parameter. Should this
shape change infinitely slowly, the evolution of the system would be
reversible. However, at any finite rate of change, it is for all
practical purposes irreversible. A simple example should suffice. 
Assume that we have a pump, inside of which the motion is chaotic. In
the pump there is one single particle, bouncing elastically off the wall
and off the piston. At some point in time someone starts moving the
piston, eventually bringing it back to its initial position. We assume
that during the entire time the piston is in motion, the person is not
permitted/allowed or able to get any information about the position or
momentum of the particle. Further, the person is not allowed to act in
any direct way on the particle. For almost all closed trajectories of
the piston, covered with a finite velocity, the particle inside will not
return to its initial state. One can then classify the ``state
transformation'' of the particle in the pump as irreversible. This in
some sense paraphrases the almost century old argument between Boltzmann
on the one side and Zermelo and Loschmidt on the other
\cite{boltzmann}. In that argument the role of the particle inside the
pump was played by the infinite number of atoms in a gas. Part of the
Boltzmann's argument was that because their number is infinite, one
cannot in any conceivable way actually reverse their velocities, and
thus irreversibility arises. We have only changed ``cannot'' with ``is
not permitted'' and arrived at the same result, but now for perhaps one
of the most simple systems possible.  This is essentially what one often
encounters in real situations. Even with the intrinsic system  having a
finite number of degrees of freedom, there is no obvious way by which
one can control their state or acquire information about their
microscopic state without disturbing them, and in particular reverse
their momenta. One can only control some ``external'' parameters. 

In Section IV we analyse a few cases of a simple quantum system, coupled
to such a complex, but finite, ``environment'' within the adiabatic
approximation. In this case the energy transfer is allowed in both
directions, from one to the other subsystem and back as well. However,
if one of the subsystems has a large number of active degrees of freedom
the energy flow will most of the time occur in one direction. This is a
generic situation, encountered in numerous quantum finite many--body
systems. 

A short summary of our results and an outlook for future investigations
are presented in the last Section V.

\section{ Evolution equations }

The Hamiltonian governing the dynamics of a quantum system coupled to a
complex environment is assumed to be of the form
\begin{equation}
H(X,x) = H_0(X) + H_1(X,x) \label{totham}.
\end{equation}
We refer to $X$ as shape variables. In the ensuing formulas we shall not
display explicitly the dependence of the Hamiltonian on the intrinsic
variables $x$ of the environment, but rather discuss its matrix elements
in a fixed intrinsic basis. The part of the total Hamiltonian
(\ref{totham}) which depends on the intrinsic coordinates $H_1(X)$ is
defined as a parametric banded random matrix, whose matrix elements
depend on the ``slow'' coordinate $X$
\begin{equation}
[H_1(X)]_{ij} = [h_{0}]_{ij} + [h_{1}(X)]_{ij}. \label{ranham}
\end{equation}
$h_0$ is taken to be diagonal and defines the average density of states,
with $\langle k|h_0|l\rangle = [h_0]_{kl}=\varepsilon _k \delta _{kl}$.
We refer to these eigenstates as ``typical states'' of the intrinsic
system with an energy $\varepsilon$. In Refs.
\cite{bdk_ann,bdk_mb8,bdk_pre} we have discussed at length the reasons
why one chooses this specific form of the Hamiltonian. For an intrinsic
subsystem with a large number of degrees of freedom the average density
of states,
\begin{equation}
\rho (\varepsilon ) = 
\overline{{\rm Tr} \delta (H_1(X)-\varepsilon )},
\end{equation}
for each given shape $X$ increases sharply with energy. The overline
denotes here a procedure for extracting the smooth part of $\rho
(\varepsilon )$ as a function of energy and it amounts essentially to an
ensemble average, to be introduced below. For a many Fermion system,
$\rho (\varepsilon )$ has a roughly exponential behaviour. Recall that
$\ln \rho (\varepsilon )$ is approximately proportional to the
thermodynamic entropy of the intrinsic system, which is an extensive
quantity. The fact that the average density of states for the intrinsic
subsystem has such a behavior is a key element of the entire approach.
This is equivalent to stating that the intrinsic subsystem has a large
heat capacity and thus can play the role of a ``reservoir'', although
not necessarily ideal. In principle $\rho (\varepsilon )$ can be
$X$--dependent as well, but we shall ignore this aspect here. Without an
$X$--dependence of the average density of states, mechanical work cannot
be performed on or by the model environment we study here, and only heat
exchange is allowed. 

In the basis of the eigenstates of $h_0$, we define $h_1(X)$ as a
parameter dependent $N\times N$ real Gaussian random matrix, which is
completely specified by its first two moments
\begin{eqnarray}
\overline{[h_1(X)]_{kl}} &=&0,\nonumber\\
\overline{ [h_1(X)]_{ij}[h_1(Y)]_{kl} } &=&
[\delta _{ik}\delta _{jl}+\delta _{il}\delta_{jk}]{\cal G}_{ij}(X-Y).
\label{correlator}
\end{eqnarray}
The overline stands for the statistical average over the ensemble of
random Gaussian matrices. Even though we shall limit all the formulas in
this work to the Gaussian orthogonal ensemble (GOE), with minor changes,
the formalism is equally applicable to other Gaussian ensembles. In Eq.
(\ref{correlator}), ${\cal G}_{ij}(X-Y)$ is a ``bell shaped''
correlation function with a characteristic width $X_0$. Physically this
is simply the formal expression of the fact that intrinsic states
corresponding to vastly different values of the shape $X$ have
statistically independent compositions. The dependence on $i,j$ allows
for the description of banded matrices, where an effective number of
states $N_0\leq N$ are coupled by $h_1(X)$. It is convenient to use an
explicit parameterization of ${\cal G}_{ij}$, which explicitly
incorporates the average density of states and the bandwidth of the
statistical fluctuations\cite{brink}:
\begin{equation}
{\cal G}_{ij}(X)= \frac{\Gamma ^\downarrow }{2 \pi\sqrt{\rho
(\varepsilon _i)\rho (\varepsilon _j)}} \exp \left [ -\frac{(\varepsilon
_i -\varepsilon _j)^2}{2\kappa _0 ^2} \right ]G \left
(\frac{X}{X_0}\right ). \label{correl}
\end{equation}
Here $G(x)=G(-x)=G^*(x)\le 1$, $G(0)=1$ and $\Gamma ^\downarrow $ is the
spreading width for the intrinsic subsystem. $\kappa_0$ (linked with the
effective band width $N_0 \approx \kappa _0 \rho(\varepsilon ))$ and
$X_0$ are also characteristics of the intrinsic system. This
parameterization is consistent with the experimental evidence that in
many--body systems the spreading width $\Gamma ^\downarrow$ changes
relatively slowly with the excitation energy\cite{brink}. Moreover,
distributions of matrix elements extracted from various theoretical
many--body models show strong deviations from a pure Gaussian
distribution \cite{matrix} and are in qualitative agreement with Eq.
(\ref{correl}). A typical Born--Oppenheimer spectrum (with an average
unit level density) as a function of the shape $X$ is shown in Fig. 1 for
a particular realization of the random Hamiltonian $H_1(X)$. For each
fixed value of the shape variable $X$ the spectrum is characterized by
fluctuations very similar to GOE.

\tdf 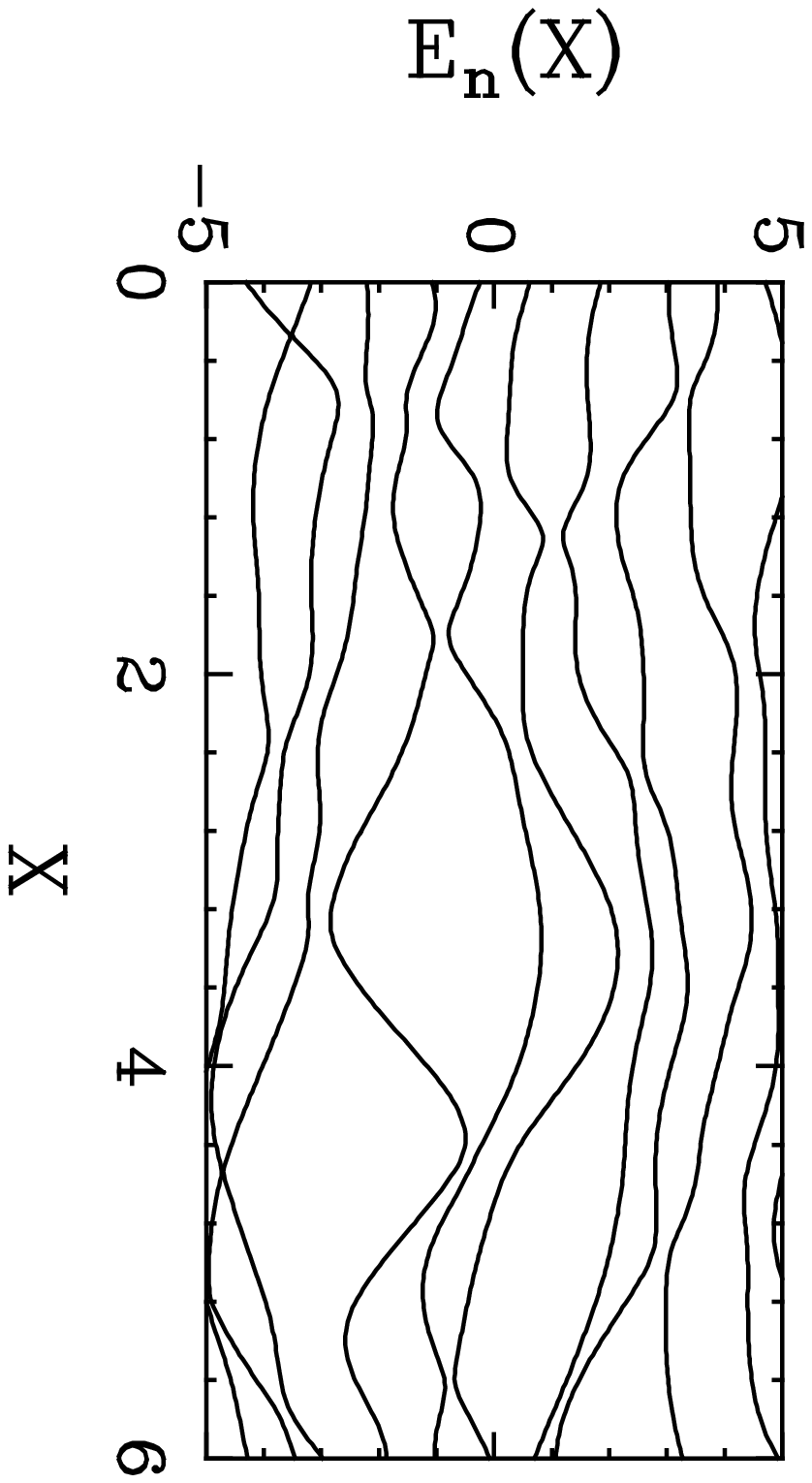 {A portion of the instantaneous eigenvalue
spectrum $E_n(X)$ as a function of the shape $X$. The Hamiltonian is
defined through Eqs. (\ref{ranham}), (\ref{correlator}) and
(\ref{correl}), with unit average density of states  and $\Gamma
^\downarrow = 2\pi$, $\kappa _0 = \infty$, $G(x)=\exp (-x^2/2)$ and
$X_0=1$.} 1. -3.5 -1. 1.

\subsection{ Driven subsystem }

A particular situation of definite physical interest is that when the
time dependence of the ``slow '' variables is known and/or controlled
externally as for example in the case of an applied external electric or/and
magnetic field(s). This is a typical situation in thermodynamics, when
the controlled parameters are changed adiabatically, insuring the
reversibility of the transformation. At any finite rate of change the
transformation looses its reversible character and dissipation sets in.
In an experiment, if one has some control over the parameters, as a
rule, one has little control on the intrinsic subsystem. Thus, if we
move the parameter $X(t)$ at finite velocity along a closed circuit in
parameter space, returning to the initial point, the final intrinsic
states will in general be distinct from the initial one. These changes
of the intrinsic subsystem, which not only depend on the rate of change
of the controlled parameters, but also on the specific paths in the
parameter space, are referred to as dissipative effects. State changes 
of the intrinsic subsystem, which depend on the path in the parameter
space, but do not depend on either the rate at which this path is
covered or on the direction in which the path is travelled, can be
incorporated into reversible type of transformations, by slightly
generalizing the definition of adiabatic transformations \cite{berry}.
Such effects can be linked with the appearance of effective abelian or
nonabelian gauge fields \cite{gauge}, besides the familiar (in the
Born--Oppenheimer approximation) effective potential forces. 

For the sake of simplicity we assume that $X(t)=V_0t$ and that the
average level density has an exponential form: $\rho (\varepsilon ) =
\rho _0 \exp( \beta \varepsilon )$. Thus $\beta =1/T=d \ln \rho
(\varepsilon )/d \varepsilon $ can be interpreted as the inverse
thermodynamic temperature of the intrinsic subsystem. (Note that the
spectrum extends to infinity in both directions from $\varepsilon =0$.)

The time evolution of the fast subsystem is found by solving the 
time-dependent Schr\"{o}dinger equation in the form
\cite{bdk_ann,bdk_mb8,bdk_pre}:
\begin{equation}
\phi (t ) = {{\rm T}}
\exp \left [-\frac{i}{\hbar } \int _0 ^{t}\! d s H_1(X(s))
 \right ]\phi (0)= {\cal {U}}(X(t))\phi (0). \label{eq:psi}
\end{equation}
where ${\rm T}$ is the time ordering operator, and ${\cal {U}}(X(t))$
the propagator. (We assume that the initial state $\phi(0)$ is
uncorrelated with the Hamiltonian $H_1(X(t))$ at later times; correlated
initial conditions have been discussed elsewhere \cite{bdk_ann}.) One
can show that in the leading order in an expansion in $1/N_0$ the
average propagator $U(X(t))=\overline{{\cal U}(X(t))}$ is diagonal in
the representation we have chosen. Its diagonal matrix elements have the
following form 
\begin{equation}
U_k(X(t)) = \overline{ \langle k | 
{\rm T}\exp \left [ -\frac{i}{\hbar}\int _0^t ds
H_1(X(s)) \right ] | k \rangle } =
\exp \left ( -\frac{i\varepsilon _k t}{\hbar } \right ) 
\sigma (X(t))
\end{equation}
( note that $\sigma (X(t))$ is state independent) and $\sigma (X(t))$
satisfies the following integral equation \cite{bdk_pre}: 
\begin{equation}
\sigma(X(t)) = 1 - \frac{\Gamma ^\downarrow }{\hbar }
 \int _0^{t} \! \! \! d s_1 \! \int _0^{s_1} \! \! \! d s_2
\sigma(X(s_1-s_2))\sigma(X(s_2))
P(s_1 -s_2)
 G \left ( \frac{(s_1 -s_2 )V_0}{X_0}\right ) . \label{propag}
\end{equation}
Here $P(s)$ is given by 
\begin{equation}
P(s) = P^*(-s) = \frac{\kappa _0}{\sqrt{2\pi } \hbar }
\exp \left [ -\frac{\kappa _0 ^2}{2\hbar ^2}
\left ( s + i\frac{\hbar \beta }{2} \right ) ^2 \right ] , \label{pes}
\end{equation}
when the correlator $\overline{ [h_1(X)]_{ij}[h_1(Y)]_{kl} }$ is defined
as in Eq. (\ref{correl}). In order to be able to compute averages of
observables, we need to introduce the set of generalized occupation
number probabilities
\begin{equation}
{\cal N}_{k}(X(t_1),X(t_2)) = 
\overline{\langle \phi (t_1)|k\rangle \langle k|\phi (t_2)\rangle }
=\sum_l \overline{\langle l |{\cal {U}}^{\dagger }
(X(t_1))|k\rangle \langle k|{\cal U}(X(t_2))| l\rangle } n_l(0).
 \label{eq:tkdef}
\end{equation}
Thus $n_{k}(t)\equiv {\cal N}_{k}(X(t),X(t))$ is the occupation
probability of the state $|k\rangle$. The generalized occupation number
probabilities can be determined by solving an evolution equation for 
the characteristic functional
\begin{eqnarray}
{\cal {N}}(X(t_1),X(t_2),\tau)& = & \overline{\langle \phi (t_1)|
\exp \left [ \frac{ih_0(\tau -t_1+t_2) }{\hbar } \right ]
|\phi (t_2)\rangle } \\
&=& \int d\varepsilon _k \rho (\varepsilon _k)
{\cal {N}}_k(X(t_1),X(t_2))
\exp \left [ \frac{i\varepsilon _k(\tau -t_1+t_2) }{\hbar }
 \right ]. \nonumber
\end{eqnarray}
The equal time functional ${\cal N}(X(t),X(t),\tau)$ is thus the Fourier
transform of the occupation number probabilities $n_k(t)$.  ${\cal
N}(t,t,\tau)$ is an extremely useful quantity, since it provides the
following cumulant expansion \cite{vankampen}:
\begin{equation}
{\cal {N}}(X(t),X(t),\tau ) = 
\sum _k n_k(t)
\exp \left ( \frac{i\varepsilon _k\tau }{\hbar } 
\right ) =
\exp \left [ \sum _n
\overline{\langle \!\langle \phi (t) |h_0^n |\phi (t)\rangle \!\rangle }
\frac{(i\tau )^n}{\hbar ^n n ! } \right ], \label{cumulant}
\end{equation}
where $\overline{\langle \!\langle \phi (t)| h_0^n |\phi (t) \rangle
\!\rangle} $ are the cumulants. If we assume that the initial occupation
number probabilities are $n_0(0)=1$ and $n_l(0)=n_{-l}(0)=0$ for $l\not=
0$ (remember that the spectrum of $H_1(X)$ is infinite in both
directions), then we find that ${\cal {N}}(X(t_1),X(t_2),\tau)$
satisfies the evolution equation \cite{bdk_pre}
\begin{eqnarray}
{\cal {N}}(X(t_1),X(t_2),\tau )&=& \sigma ^*(X(t_1))\sigma (X(t_2))+
\frac{\Gamma ^\downarrow}{\hbar }
\int _0^{t_1}\! d s_1\! \int _0^{t_2}\! d s_2
{\cal {N}}(X(s_1),X(s_2),\tau ) \label{genocc} \\
& &\times P(s_1 - s_2 - \tau )G \left ( \frac{(s_1 -s_2 )V_0}{X_0}\right
)
\sigma ^*(X(t_1-s_1))\sigma(X(t_2-s_2)). \nonumber 
\end{eqnarray}

\subsection{ The path integral approach }

When the shape variables $X$ become dynamical variables, there is energy
exchange between the two subsystems, and the dynamics becomes  more
complicated. A formalism to tackle the case when both subsystems have to
be treated quantum mechanically has been put forward by Feynman and
Vernon \cite{feynman}. One can write the following double path integral
representation for the density matrix of the entire system
\begin{eqnarray}
{\cal{R}}(X,x,Y,y,t) & = & \int d X_0 dY_0
\psi (X_0) \psi ^*(Y_0)
\int _{X(0)=X_0} ^{X(t)=X} {\cal{D}}X(t)
\int _{Y(0)=Y_0} ^{Y(t)=Y} {\cal{D}}Y(t) \\
& \times & \exp \left \{ \frac{i}{\hbar }
\left [ S_0(X(t)) - S_0(Y (t)) \right ] \right \} \nonumber \\
 & \times & \overline{
\langle x |{\rm T} \exp \left [
-\frac{i}{\hbar}
\int _{0}^{t} dt^\prime H_1(X(t^\prime )) \right ] | \phi \rangle
\langle \phi | {\rm T}_a \exp \left [
 \frac{i}{\hbar}
\int _{0}^{t} dt^{\prime \prime} H_1(Y(t^{\prime \prime }))
\right ] | y \rangle } \nonumber ,
\end{eqnarray}
where ${\rm T}$ and ${\rm T}_a$ represent the time ordering and
time anti--ordering operators respectively and $S_0(X(t))$ is the
classical action corresponding to the Hamiltonian $H_0(X)$. 
The particular form for the initial state wave function we have used
here, namely
\begin{equation}
\Psi (X,x)=\psi (X)\phi (x)
\end{equation}
is not unique and other choices are equally possible, for example, a
density matrix. By introducing the influence functional
\begin{equation}
{\cal{L}}(X(t),Y(t),t)=\overline{
\langle \phi | \left \{ \! {\rm T}_a \! \exp \left [
 \frac{i}{\hbar} \int _{0}^{t} dt^{\prime \prime }
H_1(Y(t^{\prime \prime })) \right ] \right \} \!
\left \{ \! {\rm T} \! \exp \left [
-\frac{i}{\hbar} \int _{0}^{t} dt ^\prime H_1(X(t^\prime ))
\right ] \right \}
| \phi \rangle } \label{influence}
\end{equation}
one readily obtains the following double path integral representation
for the density matrix of the ``slow'' subsystem
\begin{eqnarray}
\rho (X,Y,t) & = & \int d X_0 dY_0
\psi (X_0) \psi ^*(Y_0)
\int _{X(0)=X_0} ^{X(t)=X} {\cal{D}}X(t)
\int _{Y(0)=Y_0} ^{Y(t)=Y} {\cal{D}}Y(t) \nonumber \\
 & \times & \exp \left \{ \frac{i}{\hbar }
\left [ S_0(X(t)) - S_0(Y (t)) \right ] \right \}
{\cal{L}}(X(t),Y(t),t).
\end{eqnarray}
The formulation of the problem through a path integral representation
serves only as a very convenient vehicle to obtain an evolution equation
for the density matrix $\rho (X,Y,t)$. (Our usage of the same Greek
letter $\rho $ for two different quantities, average level density for
the intrinsic subsystem and density matrix for the ``slow'' subsystem,
should not lead to confusion.)

Using the formalism developed in Ref. \cite{bdk_pre} one can derive
relatively simple analytical expressions for the influence functional.
For the case of an adiabatic evolution of the slow subsystem, it was
shown in Ref. \cite{bdk_pre} that the influence functional has the
simple form 
\begin{equation}
{\cal{L}}(X(t),Y(t),t) = {\cal {N}}(X(t),Y(t),0 )=
\exp \left \{ \frac{\Gamma ^\downarrow }{\hbar }
\int _0^ t \left
[ G( X(t^\prime ),Y(t^\prime ) ) - 1 \right ] dt^\prime \right
\}. \label{adinfl}
\end{equation}

By combining the double path integral representation for the density
matrix $\rho (X,Y,t)$ with the above expression for the influence
functional in the adiabatic approximation, one easily derives that the
density matrix satisfies the following Schr\"{o}dinger--like equation
(for similar examples see Refs. \cite{caldeira})
\begin{equation}
i\hbar \partial _t \rho (X,Y,t) =
\{ H_0(X)-H_0(Y) + i\Gamma ^\downarrow [G(X,Y)-1]\} \rho (X,Y,t)
\label{evolution}
\end{equation}
with the initial condition
\begin{equation}
\rho (X,Y,0) = \psi (X)\psi ^*(Y).
\end{equation}
This equation for the density matrix describes a quantum mechanical
Markovian process and it satisfies the conditions of the Lindblad's
theorem \cite{lindblad}. Therefore the solutions of this evolution
equation with meaningful physical initial conditions can be given a
probabilistic interpretation at all subsequent times. In particular this
means that
\begin{eqnarray}
\rho (X,X,t) &\ge & 0 ,\\
\int dX \rho
(X,X,t) & \equiv & 1
\end{eqnarray}
for any $t\ge 0$.

\section{Temporal evolution of a driven system}

In this section we shall present analytical results for two limits,
adiabatic and diabatic, and numerical results for the intermediate
situation. In order to quantitatively determine whether the evolution of
the driven subsystem ($X(t)=V_0t$) is in either of these limits it is
useful to introduce two time scales: i) the characteristic time scale
for the slow motion $\tau _{slow}=X_0/V_0$ and ii) the characteristic
time scale for the fast degrees of freedom, $\tau _{fast}=\hbar /\kappa
_0 $. The adiabatic limit corresponds to $\tau _{slow}\gg \tau _{fast}$.
(We shall also assume that in the adiabatic limit the condition $\kappa
_0 \beta \ll 1 $ is also fulfilled.) The diabatic limit is obtained when
$\tau _{slow} \ll \tau _{fast}$.

\subsection{Adiabatic limit}

The generalized occupation number probabilities ${\cal{N}}(X(t),X(t),
\tau )$ can be computed by integrating Eqs. (\ref{propag}) and
(\ref{genocc}). The technical trick which is used is to replace in these
equations the quantity $P(s)$ introduced in Eq. (\ref{pes}) with an
appropriately chosen Dirac $\delta $--function \cite{bdk_pre}. In the
adiabatic limit $P(t)$ is much narrower than $G(V_0t/X_0)$. In the
leading order in the parameter $\kappa _0\beta \ll 1$ one obtains for $t
\ge |\tau |$
\begin{equation}
{\cal {N}}(X(t),X(t),\tau)=
\exp
\left \{ -\frac{\Gamma ^\downarrow }{\hbar }
\left [ 1 - G\left (\frac{\tau V_0 }{ X_0}\right )
\right ] (t-|\tau |) - 
\frac{\Gamma ^\downarrow |\tau |}{\hbar }\right \} .
\end{equation}
In the strict adiabatic limit $V_0\rightarrow 0$, $G(\tau V_0/X_0)
\rightarrow 1$, and the first term in the exponential vanishes. The
occupation numbers $n_k(t)$ reach rather quickly the asymptotic
distribution
\begin{equation}
n_k = \frac{1}{\pi } 
\frac{\frac{\Gamma ^\downarrow }{2} }{\varepsilon _k ^2 + 
\left (\frac{\Gamma ^\downarrow }{2} \right ) ^2},
\end{equation}
This Lorentzian shape is identical with the constant random matrix
theory result\cite{bdk_ann,bdk_pre}. During a time $t \approx \tau
_{fast}$, the slow variables hardly change and the dynamics of the fast
system is almost identical to the dynamics governed by a constant random
Hamiltonian. Our initial state in the middle of the spectrum, chosen as
$n_0(0)=1$, is thus spread over an energy interval $\approx \Gamma
^\downarrow $ and the distribution has a Lorentzian shape. If the
Hamiltonian is time independent, after this time there would be
essentially no further evolution of the average occupation number
probabilities. The subsequent dynamical evolution of the fast subsystem
occurs only because the Hamiltonian $H_1(X(t))$ is time dependent, and
only the subsequent time evolution of the system leads to dissipation
and entropy production in the long time limit.

One can now explicitly evaluate the cumulants $\overline{\langle
\!\langle \phi (t) | h_0^{n} | \phi (t)\rangle \!\rangle }$
\cite{vankampen}. All odd moments of $h_0$ vanish identically (since
$G(x)=G(-x)$ and thus there are only even powers of $\tau$ in the
expansion in Eq. (\ref{cumulant})\,). The reason for this is our
assumption that $\kappa _0 \beta \rightarrow 0$, which we lift below. In
the limit $t \rightarrow \infty $, all even cumulants of $h_0$ increase
linearly in time. If $G(x)=\exp (-x^2/2)$ (we shall use this form
hereafter for illustrative purposes) then in the limit
$t\rightarrow\infty$
\begin{eqnarray}
\overline{\langle \!\langle \phi (t)|h_0^{2n}|\phi (t)\rangle \!\rangle
}&=&
\frac{\Gamma ^\downarrow t }{\hbar } 
\left (\frac{\hbar V_0}{X_0} \right ) ^{2n}
\frac{(2n)!}{2^n n!},\nonumber\\
D(V_0)&=&
\left[\frac{\hbar \Gamma ^{\downarrow}}{2X_0^2}\right] V_0^2
=\left[\frac{\hbar \Gamma ^{\downarrow}}{2\tau _{slow}^2}\right],
\end{eqnarray}
resulting in a non--Gaussian distribution. A Gaussian process would have
only nonvanishing linear and quadratic cumulants. The energy diffusion
constant is extracted from the time dependence of the second cumulant
when $t\rightarrow \infty$ according to
\begin{equation}
\Delta_E^2(t)= 
\overline{ \langle \phi (t)|[H_1(X(t))-E(t)]^2| \phi (t) \rangle }
\approx 
\overline{ \langle \phi (t)|h_0^2| \phi (t) \rangle }
\approx 
\; {{\rm const}}\; + \; 2D(V_0)t , \label{diff}
\end{equation}
where $E(t)= \overline{ \langle\phi (t)|H_1(X(t))| \phi (t) \rangle }$.
Note that the energy variance is time dependent only for a time
dependent Hamiltonian. As the result of the symmetric initial
distribution and since $\beta = 0$ (remember that $\kappa _0\beta \ll 1
$), we obtain no friction (i.e. $E(t)={{\rm const}}$) and only 
nonvanishing even cumulants. To get friction we have to consider the
next order corrections to the adiabatic limit $\kappa _0\beta \ll 1 $.
Since for $\beta >0$ the average level density is increasing with
energy, there will be on the average more transitions upward in energy than
downward and thus the driven subsystem is heated up. One can show that
the odd cumulants are then given by the following expressions\cite{bdk_pre} 
\begin{equation}
\langle \!\langle \phi (t) | h_0 ^{2n-1} | \phi (t)\rangle\!\rangle
=\frac{\beta }{2} \langle \!\langle \phi (t) | h_0 ^{2n} | \phi
(t)\rangle\!\rangle .
\end{equation}
The case $n=1$ corresponds to the Einstein fluctuation--dissipation
theorem. In familar diffusive processes, all cumulants of order higher
than the second are vanishing. The existence of large higher order
cumulants results in energy tails of the energy distribution
significantly longer than in traditional phenomenological transport
approaches, like Fokker--Planck or Langevin equations. One can show that
in the tails the energy distribution has the following behaviour
\begin{equation}
P(\varepsilon
)\propto \exp ( -\alpha |\varepsilon |\ln ^{1/2} |\varepsilon| ) ,
\label{energydis}
\end{equation}
where $\alpha $ is some (time dependent) constant. A somewhat similar
functional form has been determined for the distribution of conductance
fluctuations in mesoscopic systems \cite{lerner}. The presence of these
longer than expected tails is a clear indication that the excitation
mechanism cannot be reduced to a simple random walk in energy space.

\subsection{Diabatic limit}

Another simple analytical solution can be obtained in the diabatic
limit, when $\tau _{slow} = X_0/V_0 \ll \tau _{fast} = \hbar /\kappa _0
$. In this case $G(V_0s/X_0)$ becomes much narrower than $P(s)$ and can
be replaced with an appropriately chosen Dirac $\delta $--function. The
evolution equations (\ref{propag}) and (\ref{genocc}) can be solved
again and one obtains:
\begin{equation}
{\cal {N}}(t,t,\tau )=
\exp
\left \{ -\frac{\Gamma ^\downarrow X_0 \kappa _0 }{\hbar ^2 V_0}
\left [ \exp \left( \frac{\kappa _0 ^2\beta ^2 }{8}\right ) -
\exp \left ( \frac{\kappa _0 ^2}{2}\left (\frac{\beta }{2}+
\frac{i\tau }{\hbar} \right ) ^2 \right )
 \right ] t \right \} .
\end{equation}
(Note that in this case what we denote as the slow degrees of freedom
are actually faster than the intrinsic ones.) This is similar to the
functional form found in the adiabatic limit. Again, all the cumulants
of $h_0$ increase linearly in time
\begin{equation}
\overline{\langle \!\langle \phi (t)|h_0^n|\phi (t)\rangle \!\rangle }
=\left [ \frac{\Gamma ^\downarrow X_0 \kappa _0 }{\hbar ^2V_0} \exp
\left ( \frac{\beta ^2\kappa _0 ^2}{8}\right ) \left ( \frac{i\kappa _0
}{\sqrt{2}} \right )^n {{\rm H}}_n\left ( -\frac{i\kappa _0 \beta
}{2\sqrt{2}} \right ) \right ]\; t,
\end{equation}
where ${{\rm H}}_n(x)$ are Hermite polynomials, resulting in a
non--Gaussian diffusion of the occupation numbers. From Eq. (\ref{diff})
and the second cumulant, we find in this limit a completely different
velocity dependence:
\begin{equation}
D(V_0) = \left[\frac{\Gamma^\downarrow
X_0\kappa_0^3(\beta^2\kappa_0^2+4)}
 {8\hbar^2}
\exp\left(\frac{\beta^2\kappa_0^2}{8}\right)\right]\frac{1}{V_0}.
\end{equation}

\tdf 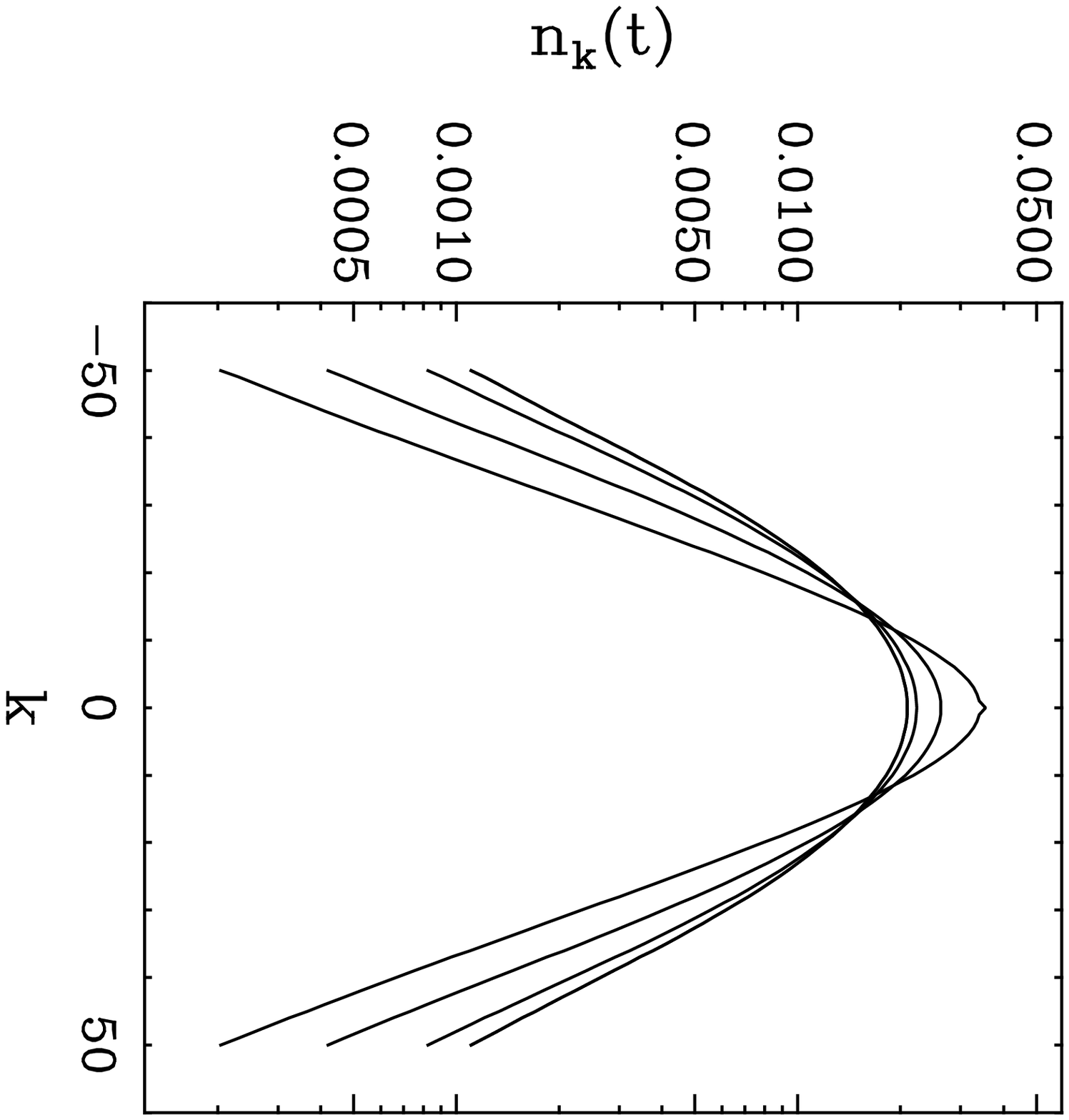 {  The time dependence of the occupation
probabilities $n_k(t)$ (note the vertical logarithmic scale) plotted as
functions of the level number $k$, for $V_0=4$, $\beta =0$ and $\kappa
_0 \approx 20$. The values of the remaining parameters are specified in
the text. The narrower distribution corresponds to $t=1$ with widening
curves for $t=2$, 3 and 3.5 respectively. The small notch in the
uppermost curve is a remnant of the initial conditions $n_0(0)=1$ and
$n_{k\ne 0}(0)=0$.} 0. 1. -1.2 1.

\noindent
The fluctuation--dissipation theorem, which provides the relation
between the first and the second cumulants is in this case:
\begin{equation}
\beta D= \gamma \left ( 1+\frac{\beta ^2\kappa _0 ^2}{4}\right ),
\end{equation}

\tdf 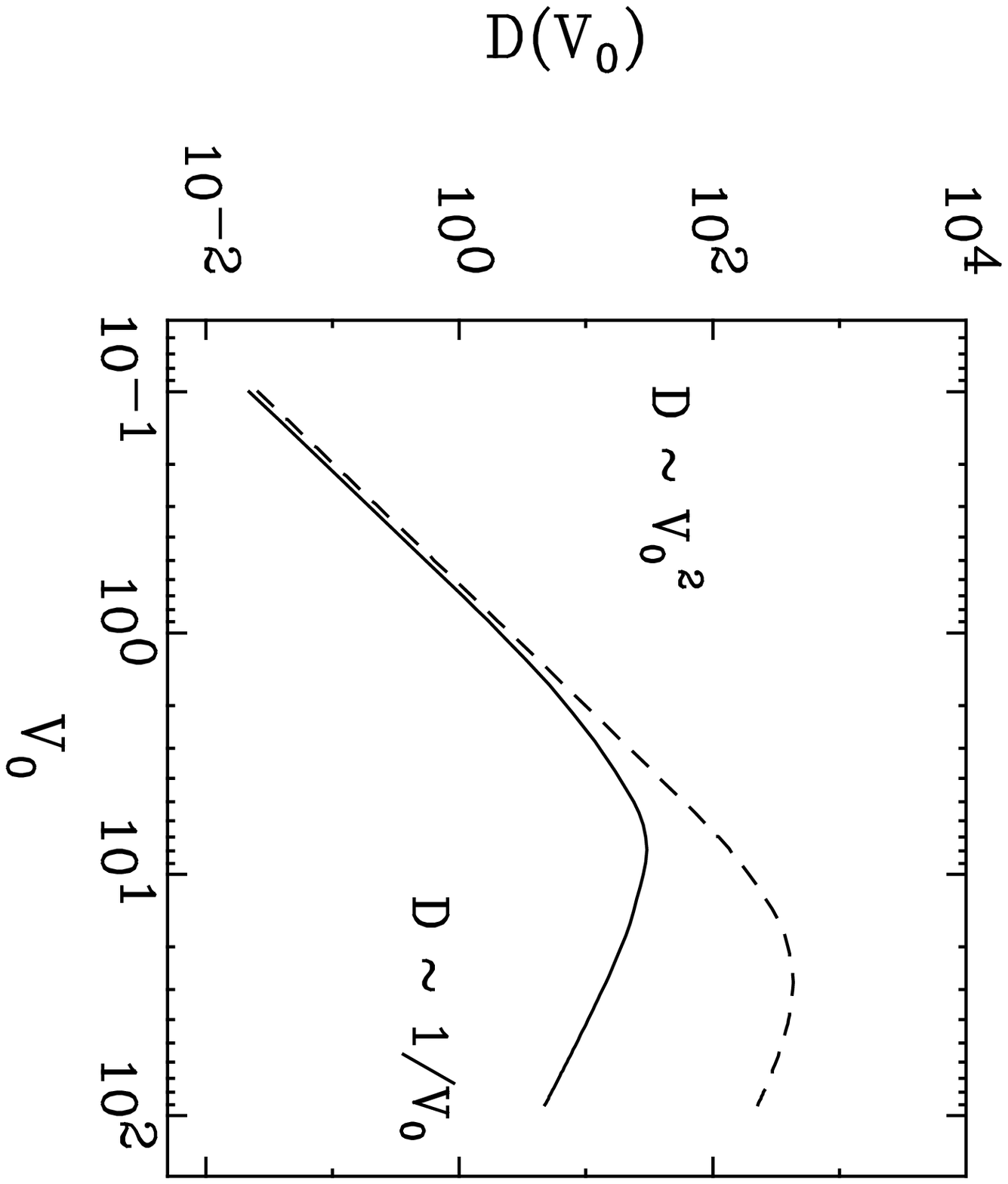 {  Velocity dependence of the diffusion constant
$D(V_0)$ from the adiabatic to the diabatic limits. We use $X_0=1$,
$\beta /\rho _0 = 0.1$,  $\rho_0 \Gamma ^\downarrow =2\pi$ ($\rho _0$
thus defines the energy units), $\kappa_0\rho_0=5$ (the lowest curve)
and $\kappa_0 \rho_0=15$ (the highest curve). } 0.1 -1.3 -1. 1.

\noindent where $\gamma t = \overline{\langle \!\langle \phi (t)|h_0|\phi
(t)\rangle \!\rangle }$. For a Gaussian diffusive process one would
have obtained $\beta D= \gamma $. Thus, in both adiabatic and diabatic
limits we have obtained significant deviations from an excitation
mechanism corresponding to a simple random walk in the energy space. One
might expect that the intermediate regime will be very similar in this
respect, as we shall substantiate in the next subsection. We thus
conclude that the energy transfer to a ``complex'' quantum system is
definitely not a Markovian process. The presence of the 
above mentioned
long tails shows that the system retains for some time the memory of the
direction it was proceeding and transitions into the same direction in
energy are somewhat favored over transitions into the other direction.

\subsection{Arbitrary driving velocity}

For the intermediate velocity regime we have to resort to a numerical
solution of the evolution equations Eqs. (\ref{propag}) and
(\ref{genocc}). In Fig. 2 we show $n_k(t)$ as function of the level
number $k$. The tails have an almost pure exponential behavior, in
agreement with the theoretical expectations, see Eq. (\ref{energydis}).
In Fig. 3 we show the behavior of the diffusion constant, $D(V_0)$, from
the adiabatic to the diabatic limit for some values of the parameter
$\beta$. In all cases, $D(V_0)$ evolves from quadratic (in the adiabatic
limit) to an inverse velocity dependence (in the diabatic limit). At
high velocities, the system becomes increasingly transparent, as
reflected in the decrease of the diffusion constant. A similar behaviour
is observed for the first cumulant, i.e. the average heating rate, which
as a function of the velocity $V_0$ has a similar profile with $D(V_0)$.
This is reminiscent of the motional narrowing phenomenon in NMR. At low
enough velocities the fast system has sufficient time to
``accommodate'' to the new environment, while the shape $X$ changes. In
the opposite limit of high velocities, the shape $X$ evolves so rapidly
that the system can barely react to the changes. Consequently, the
energy diffusion is maximal only for some intermediate velocity regime,
when the ``slow'' motion is in ``resonance'' with the ``fast'' dynamics,
namely when $\tau _{slow}=X_0/V_0$ is comparable to $\tau _{fast}=\hbar
/\kappa _0$. In Fig. 4, we plot the ratio of $\beta D/\gamma$, where $D$
and $\gamma$ are computed from the first and second order cumulants as a
function of velocity $V_0$. When the ratio is unity, the Einstein limit
of the fluctuation--dissipation theorem is recovered. Noticeable
differences occur at large velocities. Higher order cumulants have also
been extracted from our numerical results and in all cases their
magnitudes and temporal behavior is similar to the analytical results
discussed in the previous two subsections. Higher order cumulants
increase essentially linearly with time and their magnitudes increase
with the order of the cumulant.

\tdfullfigure 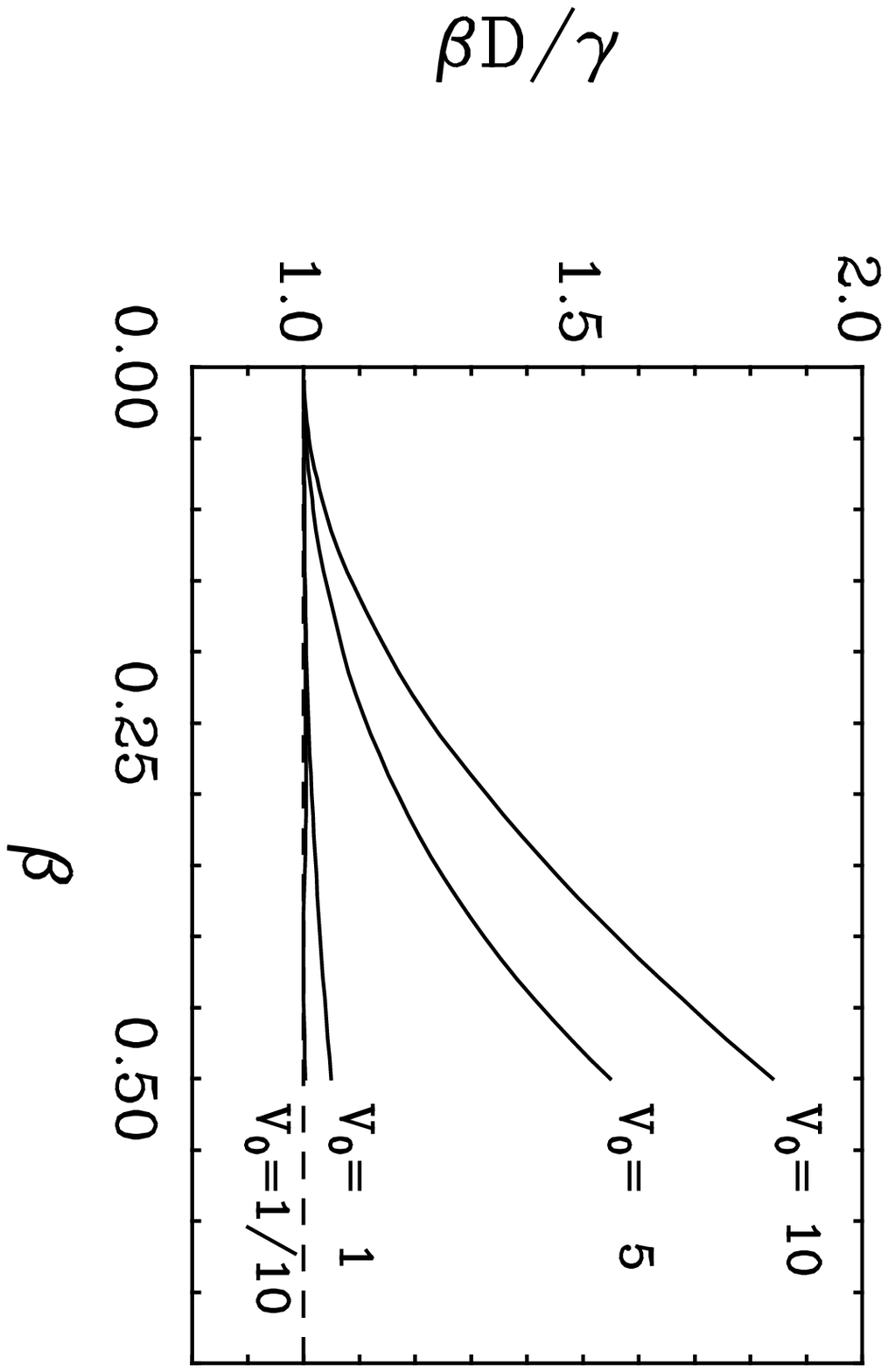 {  Deviation from the fluctuation--dissipation
theorem as a function of $\beta=1/T$. For low velocities, the theorem is
largely satisfied. However, for large velocities it is violated even at
moderate temperatures. Here we use $X_0=1$, $\rho _0\Gamma ^\downarrow
=2\pi $ and $\kappa_0\rho _0=5$.} 0.3 0. -.5 1
                                     
\section{Dynamical evolution of a simple quantum system in a
complex environment}

In the adiabatic approximation (discussed in the previous section), it is
possible to analytically determine the influence functional.
Since the characteristic time scale for the intrinsic subsystem $\tau
_{fast}=\hbar /\kappa _0 $ is much shorter than any expected time scale
of the ``slow'' subsystem in Eq. (\ref{influence}) one obtains
significant contributions only from ``left '' $Y(t^{\prime \prime})$ and
``right'' $X(t^\prime )$ paths corresponding to $t^{\prime \prime }
=t^\prime $, and the influence functional acquires the particularly
simple form given in Eq. (\ref{adinfl}). For a derivation of this form
we refer the reader to Ref. \cite{bdk_pre}. As a result the evolution equation
for the density matrix of the ``slow'' subsystem is local in time and
memory effects are absent. Among the conditions we have listed for
the applicability of our adiabatic results, we had $\beta =0$,
which corresponds to an infinite temperature of the
intrinsic subsystem (since $\beta = 1/T$). As we have discussed in
Section II, in the present formulation only heat transfer is allowed
between the two subsystems. Since the temperature of the intrinsic
system is infinite in this limit, one naturally expects energy transfer
only from the intrinsic system towards the slow system and no mechanical
work. Even though there are no memory effects in the time evolution of
the ``slow'' subsystem, memory effects are still present in the
evolution of the intrinsic subsystem (see Section III).

It is worth noting that the functional form of the
influence functional derived by us is entirely different from the
popular Caldeira--Leggett form \cite{caldeira}, which is a quadratic
expression in $X(t)$ and $Y(t)$. If we were to use only the first term
in a Taylor expansion of $G( X(t^\prime ),Y(t^\prime ) )-1$ we would
obtain an expression similar to Caldeira--Leggett form for the influence
functional. 

It is well known that most of the trajectories in a path integral
are very jagged and one might question the applicability of an adiabatic
approximation. One can try to make a long argument in 
support of the adiabatic approximation using 
``circumstantial'' evidence, e.g. the success of the
Born--Oppenheimer approximation in studying various physical systems. We
only note that relatively smooth paths, obtained by taking into account
only low frequencies, seem to be adequate, as is clearly shown in the
partial Fourier smoothing methods for computing path integrals
\cite{fourier}. 

In this section we shall discuss several cases: that of a linear
potential, a quadratic potential and a double well potential where
dissipative tunneling can occur. Surprisingly enough, the first two
cases can be solved explicitly. To our knowledge,
these represent new cases of time dependent quantum mechanical problems,
which can be given an entirely analytical treatment \cite{baz}. Only the
case of a double well potential is treated numerically.

\subsection{Linear potential}

The case when the slow variables evolve in a linear potential
\begin{equation}
H_0(X)=-\frac{\hbar ^2}{2m}\partial _X^2 -FX
\end{equation}
is particularly instructive. We shall assume furthermore that the
``effective potential'' is ``translation invariant'', namely
$G(X,Y)=G(X-Y)$. In terms of the variables $s=X-Y$ and $r=(X+Y)/2$ the
evolution equation for $\rho (X,Y,t)$ becomes
\begin{equation}
(i\hbar \partial _t + \frac{\hbar ^2}{m}\partial _r \partial _s )
\rho (r,s,t) = \{-Fs + i \Gamma^\downarrow [ G (s ) - 1 ]\} \rho (r,s,t)
\label{eveq}.
\end{equation}
We seek a solution in the form
\begin{equation}
\rho (r,s,t)= \int \frac{d k}{2\pi \hbar}
\exp \left ( \frac{ikr}{\hbar} \right ) d(k,s,t)
\label{dtran}.
\end{equation}
The function $d(s,t,k)$ satisfies the equation
\begin{equation}
\left ( \partial _t + \frac{k}{m}\partial _s \right ) d(k,s,t) =
\left \{ \frac{iFs}{\hbar }+ \frac{\Gamma^\downarrow }{\hbar }
[G(s)-1]\right \} d(k,s,t) . \label{deveq}
\end{equation}
For either $s=0$ or $k=0$, $d(k,s,t)$ is the characteristic function
\cite{vankampen} for the spatial or momentum distribution of the slow
subsystem, respectively.

Using the method of characteristics for wave equations \cite{whitham},\,
Eq. (\ref{deveq}) can be solved through quadratures and the density
matrix is determined to be 
\begin{eqnarray}
\rho (r,s,t) &=& \int \!\!\int \frac{dr^\prime d k}{2\pi \hbar }
\rho _0\left ( r^\prime , s-\frac{kt}{m} \right )
\exp \left [ \frac{ik(r-r^\prime)}{\hbar} \right ]\nonumber \\
& \times &\exp \left \{ \frac{iFst}{\hbar}
- \frac{iFt^2k}{2\hbar m} +
\frac{\Gamma^\downarrow m}{\hbar k}
\int _{ s-\frac{kt}{m}}^{s} ds^\prime [G(s^\prime )-1] \right \} .
\label{linden}
\end{eqnarray}
where $\rho _0 (r,s) =\rho (r,s,0)$ is the initial density matrix. From
Eqs. (\ref{dtran}) and (\ref{linden}), the characteristic function for
the momentum distribution can be identified as $D(s,t)= d(0,s,t)$,
where:
\begin{equation}
D(s,t)=\int dr \rho (r,s,t) = \int dr
\rho _0( r,s) \exp \left \{
\frac{iFst}{\hbar } +
\frac{\Gamma^\downarrow t}{\hbar }[G(s)-1] \right \} .
\end{equation}
One extremely economical and intuitive way to characterize the momentum
distribution of the collective subsystem is through its
cumulants
\begin{equation}
\langle \!\langle p^n\rangle \!\rangle |_t
= \left . \left ( \frac{ \hbar d }{i ds}\right ) ^n
 \ln D(s,t) \right | _{s=0} 
\end{equation}
For the case of a Gaussian correlation function $G(X)=\exp [
-X^2/2X_0^2]$, we find
\begin{eqnarray}
\langle \!\langle p \rangle \!\rangle |_t &=&
\langle \!\langle p \rangle \!\rangle |_{t=0} + Ft , \\
\langle \!\langle p^{2} \rangle \!\rangle |_t &=&
\langle \!\langle p^{2} \rangle \!\rangle |_{t=0} +
\frac{\Gamma^\downarrow \hbar t}{X_0^{2}} ,\\
\langle \!\langle p^{2n} \rangle \!\rangle |_t &=&
\langle \!\langle p^{2n} \rangle \!\rangle |_{t=0} +
(2n-1)!! \frac{\Gamma^\downarrow t}{\hbar } \;
\left ( \frac{ \hbar }{X_0}\right ) ^{2n} \\
\langle \!\langle p^{2n+1} \rangle \!\rangle |_t &=&
\langle \!\langle p^{2n+1} \rangle \!\rangle |_{t=0}.
\end{eqnarray}
The meaning of the ``correlation length'' $X_0$ is that intrinsic shapes
separated by $|X-Y|> X_0$ are statistically uncorrelated. Notice that
only the first cumulant is affected by the presence of a linear
potential in the expected manner, namely a uniform acceleration of the
slow subsystem. The ``bath'' of intrinsic degrees of freedom affects
only higher order even cumulants of the momentum distribution while the
odd cumulants of order higher than one remain unchanged. 

The cumulants of the spatial distribution can be obtained from the
characteristic function $d(k,0,t)$. From Eq. (\ref{linden}) it
immediately follows that
\begin{eqnarray}
\sum _{n=0}^{\infty } \frac{1}{n!}\left ( -\frac{ik}{\hbar } \right ) ^n
\langle \!\langle r^n \rangle \!\rangle &=&
\ln d_0(k,0,t) + \left ( - \frac{ik}{\hbar }\right ) \frac{Ft^2}{2m}
 \nonumber\\
&+& \frac{\Gamma^\downarrow m}{\hbar k}\int _{-\frac{kt}{m}}^0
ds^\prime [G(s^\prime )-1] .
\label{xcums}
\end{eqnarray}
The term $\ln d_0(k,0,t)$ gives the contributions to the cumulant
expansion arising from free expansion of the initial wave
packet, in the absence of both the linear potential and the coupling to
the internal degrees of freedom. The linear potential leads to the
expected (classical) behaviour of the center of the wave packet (see the
second term on the rhs of Eq. (\ref{xcums})\,). The contribution to the even
cumulants arising from dissipation alone is
\begin{equation}
\langle \!\langle r^{2n} \rangle \!\rangle |_{diss}=
\frac{(2n-1)!!}{2n+1}\;
\frac{\Gamma^\downarrow t}{\hbar }\;
\left ( \frac{ \hbar t}{m X_0} \right ) ^{2n}.
\end{equation}
Of particular interest is the second cumulant
\begin{equation}
\langle \!\langle r^{2} \rangle \!\rangle |_{diss}=
\frac{\Gamma^\downarrow \hbar t^3 }{3X_0^2m^2},
\end{equation}
which shows that dissipation leads to a super diffusive expansion of the wave
packet. This behaviour is to be contrasted with the free expansion or
ballistic propagation, in which case $\langle \!\langle r^{2} \rangle \!
\rangle \propto t^2$ and with normal diffusion, for which $\langle
\!\langle r^{2} \rangle \!\rangle \propto t$. It is instructive to
estimate also the ``size'' of this state in phase space. In the
limit $t\rightarrow \infty$ the dissipative contribution dominates,
and one has
\begin{equation}
\Delta r \Delta p \approx \left [
\langle \!\langle r^{2} \rangle \!\rangle |_{diss}
\langle \!\langle p^{2} \rangle \!\rangle |_{diss} \right ] ^{1/2}=
\frac{2\pi W_0 \hbar t^2}{\sqrt{3}mX_0^2} .
\end{equation}
Thus the ``blob'' spreads in phase space at a much faster rate than
in the case of a simple quantum mechanical system.

Similarly, one can show that with logarithmic accuracy for large values
of the variable $\theta (r,t)=|r-Ft^2/2m|/t$ (assuming vanishing initial
average linear momentum) the spatial distribution behaves as
\begin{equation}
\rho(r,0,t) \propto 
\exp \left [ -\nu \theta (r,t) \ln ^{1/2} \theta (r,t) \right ] ,
\label{uncert}
\end{equation}
where $\nu $ is some constant. Thus the effect of dissipation is
undeniably not only significant, but also leads to qualitatively new
features.

\subsection{Quadratic potential}

Another case that is susceptible of an analytical treatment is that of 
a quadratic potential for the collective subsystem
\begin{equation}
H_0(X)=-\frac{\hbar ^2}{2m}\partial _X^2 + \frac{m\omega ^2X^2}{2} .
\end{equation}
Using the representation for $\rho (r,s,t)$ of Eq. (\ref{dtran}), the
equation for the transformed density now becomes
\begin{equation}
\left ( \partial _t + \frac{k}{m}\partial _s - m\omega ^2 s
\partial _k \right ) d(k,s,t) =
\frac{\Gamma^\downarrow }{\hbar }
[G(s)-1]d(k,s,t).
\end{equation}
The method of characteristics~\cite{whitham} can again be used to
determine its solution 
\begin{eqnarray}
\rho (r,s,t) &=& \int \frac{d k}{2\pi \hbar} 
\exp \left [ \frac{ikr}{\hbar }\right ] \nonumber\\
&\times& d_0\left ( s \cos \omega t-\frac{k}{m\omega }\sin \omega t ,
m\omega s \sin \omega t + k \cos \omega t \right ) \\
& \times & \exp \left \{ \frac{\Gamma^\downarrow }{\hbar }
\int _0^t dt^\prime
[G(s\cos \omega (t-t^\prime )
-\frac{k}{m\omega }\sin \omega (t-t^\prime ) )-1]
\right \} ,\nonumber
\end{eqnarray}
where $d_0(s,k)=d(s,k,0)$. In a similar manner to the one described in
the previous subsection, one can determine various cumulants. For both
spatial and momentum distributions only even cumulants are affected by 
dissipation \cite{ryzhik}
\begin{eqnarray}
\langle \!\langle p^{2n} \rangle \!\rangle |_{diss} &=&
(2n-1)!! \frac{\Gamma^\downarrow }{\hbar \omega } \;
\left ( \frac{\hbar}{X_0} \right )^{2n}
\int _0 ^{\omega t} d \tau \cos ^{2n} \tau \\
& \approx & \frac{ [(2n-1)!!]^2}{2^n n!} \; 
\frac{ \Gamma^\downarrow t}{\hbar } \;
\left ( \frac{\hbar}{X_0} \right )^{2n}, \\
\langle \!\langle r^{2n} \rangle \!\rangle |_{diss} &= &
(2n-1)!! \frac{\Gamma^\downarrow }{\hbar \omega } \;
\left ( \frac{\hbar}{m\omega X_0} \right )^{2n}
\int _0 ^{\omega t} d \tau \sin ^{2n} \tau \\
 & \approx & \frac{ [(2n-1)!!]^2}{2^n n!} \; 
\frac{\Gamma^\downarrow t}{\hbar } \;
\left ( \frac{\hbar}{m\omega X_0} \right )^{2n}.
\end{eqnarray}
There is a noticeable difference with the case of a linear potential, in
that all cumulants increase now only linearly with time. It looks as if
the quadratic potential has a ``focusing'' effect on the spatial
distribution. The fact that the momentum and spatial distributions are
so similar should come as no surprise in the case of a harmonic
oscillator, which possesses an obvious symmetry between the momenta and
coordinates. In the limit $t\rightarrow \infty$ we obtain that the
``uncertainty relation'' has the following expression
\begin{equation}
\Delta r \Delta p \approx \left [
\langle \!\langle r^{2} \rangle \!\rangle |_{diss}
\langle \!\langle p^{2} \rangle \!\rangle |_{diss} \right ] ^{1/2}=
\frac{\pi \hbar W_0 t}{m\omega X_0^2} ,
\end{equation}
which has to be contrasted with the quadratic time behaviour in the
linear or no potential cases, see Rel. (\ref{uncert}).

It is a simple matter to analytically continue these expression to the
case of an inverted parabolic potential or barrier. This rather
innocuous procedure, leads however to an entirely different time
dependence of the cumulants, all of them increasing exponentially with
time in this case (as $\cos \tau$ and $\sin \tau $ become $\cosh \tau$
and $\sinh \tau$ respectively).

\subsection{ Tunneling in a symmetric double well potential } 

The double well potential we analyse has the form
\begin{equation}
V(X) = a \left ( X^2-\frac{b}{2a}\right )^2. \label{doublewell}
\end{equation}
For a strong enough barrier the spectrum of a relatively large number of
low lying eigenstates is made up of doublets in the absence of the
coupling to the intrinsic subsystem. The corresponding eigenfunctions
are approximately symmetric and antisymmetric combinations of wave
functions localized in the two wells. The first nine eigenvalues are
shown in Fig. 5 for the case $a=1/2$ and $b=5$ together with the
potential. We have chosen as an initial state a linear combination of
the eigenfunctions of the ground and first excited states ($n=0,1$)
\begin{eqnarray}
H_0(X)&=& -\frac{\hbar ^2}{2m}\partial _X^2 + V(X),\\
H_0(X)\psi _n(X) &=& E_n \psi _n(X)\\
\rho(X,Y,0)&=& \frac{1}{2}[\psi _0(X)-\psi _1(X)][\psi _0(Y)-\psi
_1(Y)]. \label{initial}
\end{eqnarray}
In this case the particle is initially localized in one well with a 
probability almost equal to one, and only an exponentially small amount
is present
in the other well. In the absence of coupling to the intrinsic subsystem
this state will tunnel almost entirely to the other well in an
exponentially long time $\tau _{tunnel} = \pi \hbar /(E_1-E_0)$, since
the splitting between the two states is exponentially small. For the
particular choice of parameters we have chosen ($\hbar =1$, $m=1/2$,
$a=1/2$ and $b=5$) $\tau _{tunnel} \approx 2,300$. 

\tdfullfigure 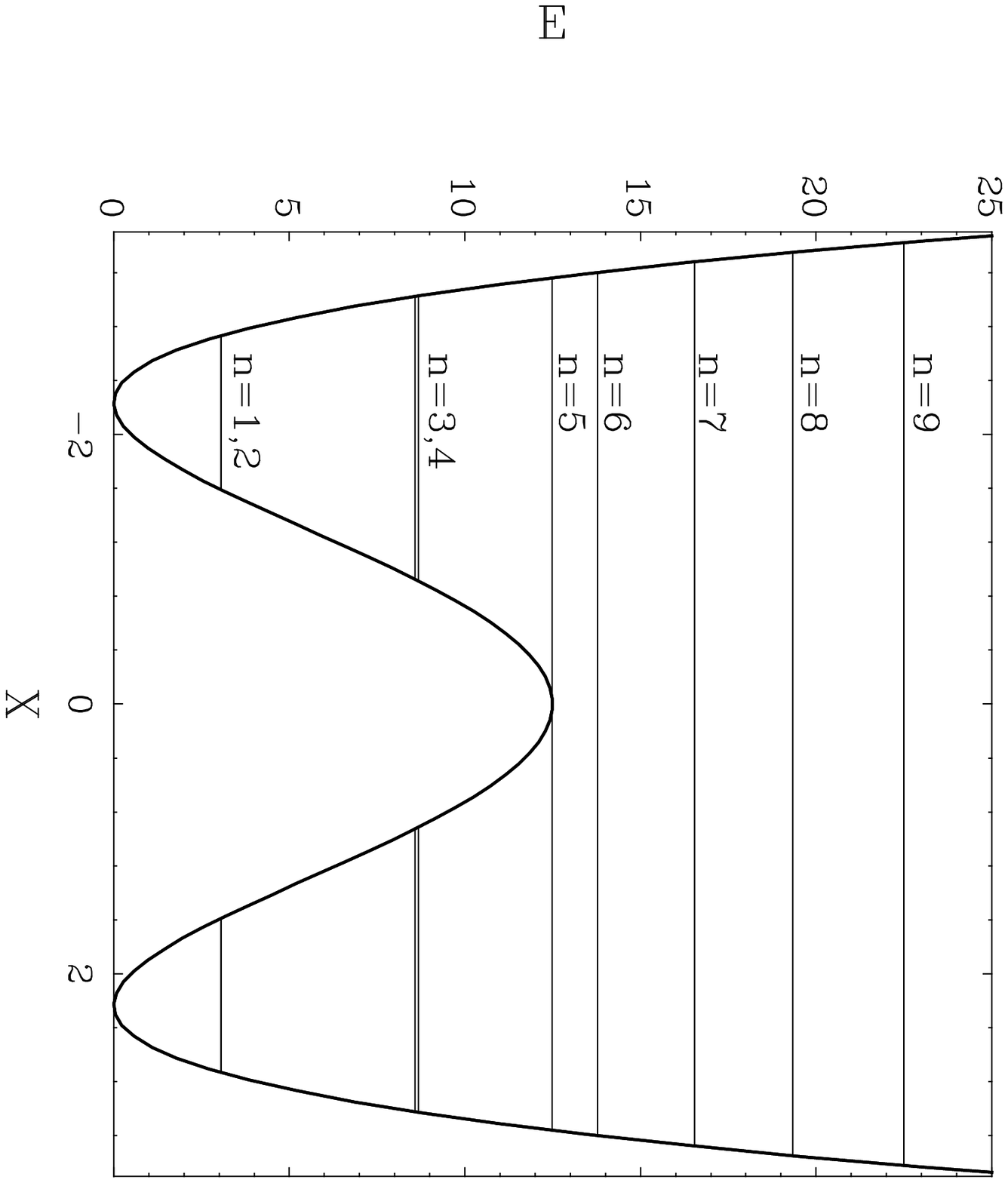
{  The first nine eigenvalues for the Schr\"odinger
equation with the potential (\ref{doublewell}) with $\hbar=1$, $m=1/2$,
$a=1/2$ and $b=5$. (The splitting of the first two levels is
approximately $10^{-3}$.)} 0.5 4 -.5 1

The coupling to the intrinsic subsystem is characterized by two
parameters $\Gamma ^\downarrow$ and $X_0$, which we have varied
independently. We retain a Gaussian correlator $G(x)=\exp (-x^2/2)$. As
we show in Figs. 6 and 7, the effect of dissipation on the tunneling
process is profound. Not only the tunneling rate is changed by orders of
magnitude, but the shape of the wave packet is qualitatively different
from the one in the absence of coupling. In the case of an usual quantum
mechanical tunneling, the wave packet would have simply gradually
changed from a state localized in one well to a state in the other well.
Since the potential well is symmetric, the shape of the wave packet
after a time $\tau _{tunnel} = \pi \hbar /(E_1-E_0)$ would have been
simply the mirror image with respect to the origin of the initial one,
see Fig. 6.

\tdfullfigure 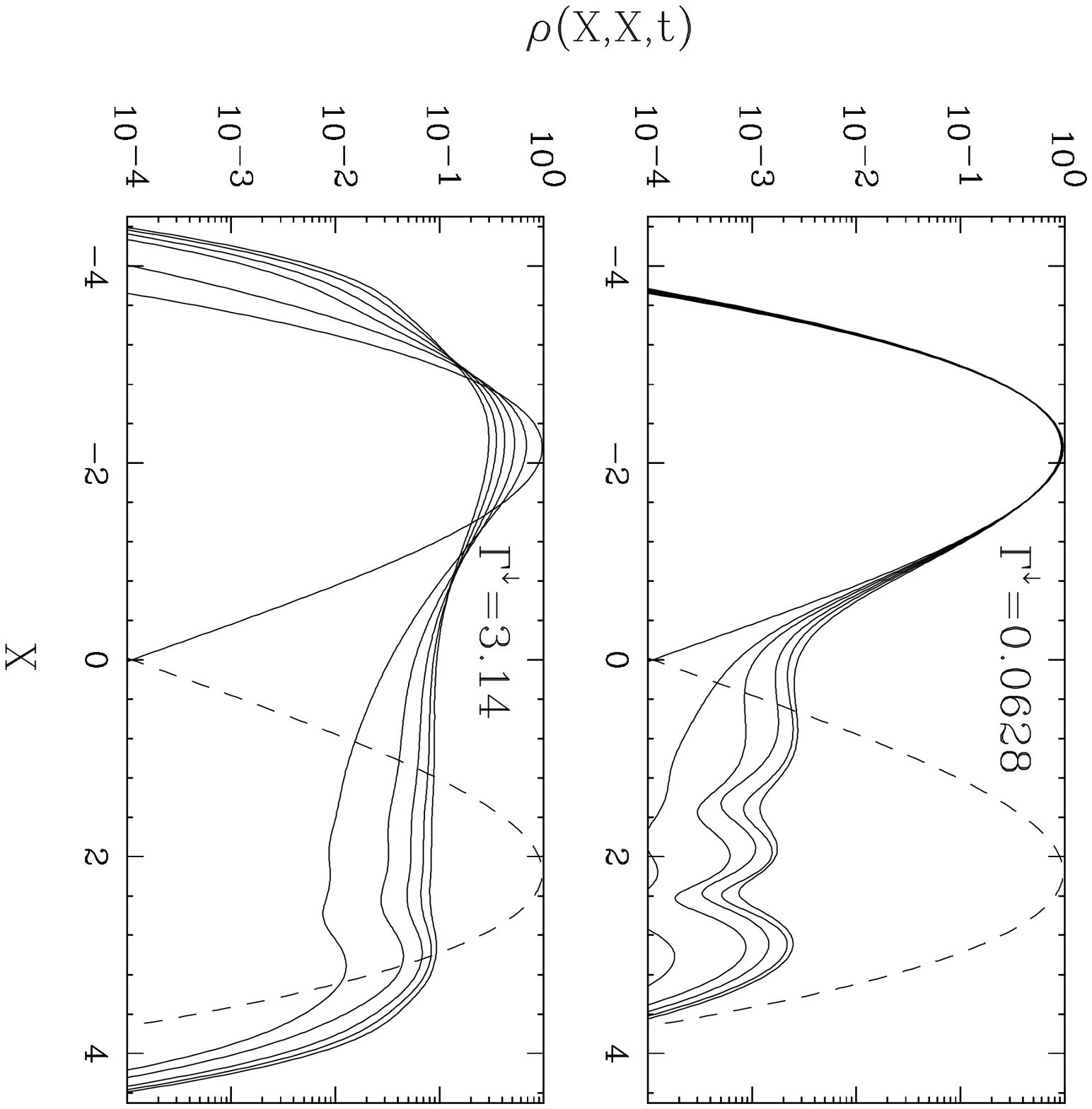
{  The density profiles for $t=0$, 1, 2, 3, 4 and 5
for the initial condition defined in Eq. (\ref{initial}), in the double
well potential of Fig. 5 for two different values of $\Gamma
^\downarrow$. The dashed line shows the density profile in the absence
of dissipation after a time $t=\tau _{tunnel} = \pi \hbar
/(E_1-E_0)\approx 2,300$. } 0.2 6 -.5 1

The role of the spreading width $\Gamma ^\downarrow$ is rather simple to
understand. Since the coupling between the two subsystems is defined by
it, see Eq. (\ref{correl}), it is natural to expect that with increasing
$\Gamma ^\downarrow$ the rate of tunneling increases as well. We observe
an approximate power law relationship between the probability to find
the particle in the other well at a given time 
\begin{equation}
P_+(t) =\int _0 ^\infty dX \rho (X,X,t) 
\end{equation}
and the spreading width $\Gamma ^\downarrow$ (see Fig. 7), namely
$P_+(t)\propto {\Gamma ^\downarrow }^{\alpha }$, where $\alpha \approx
3/2$.

\tdfullfigure 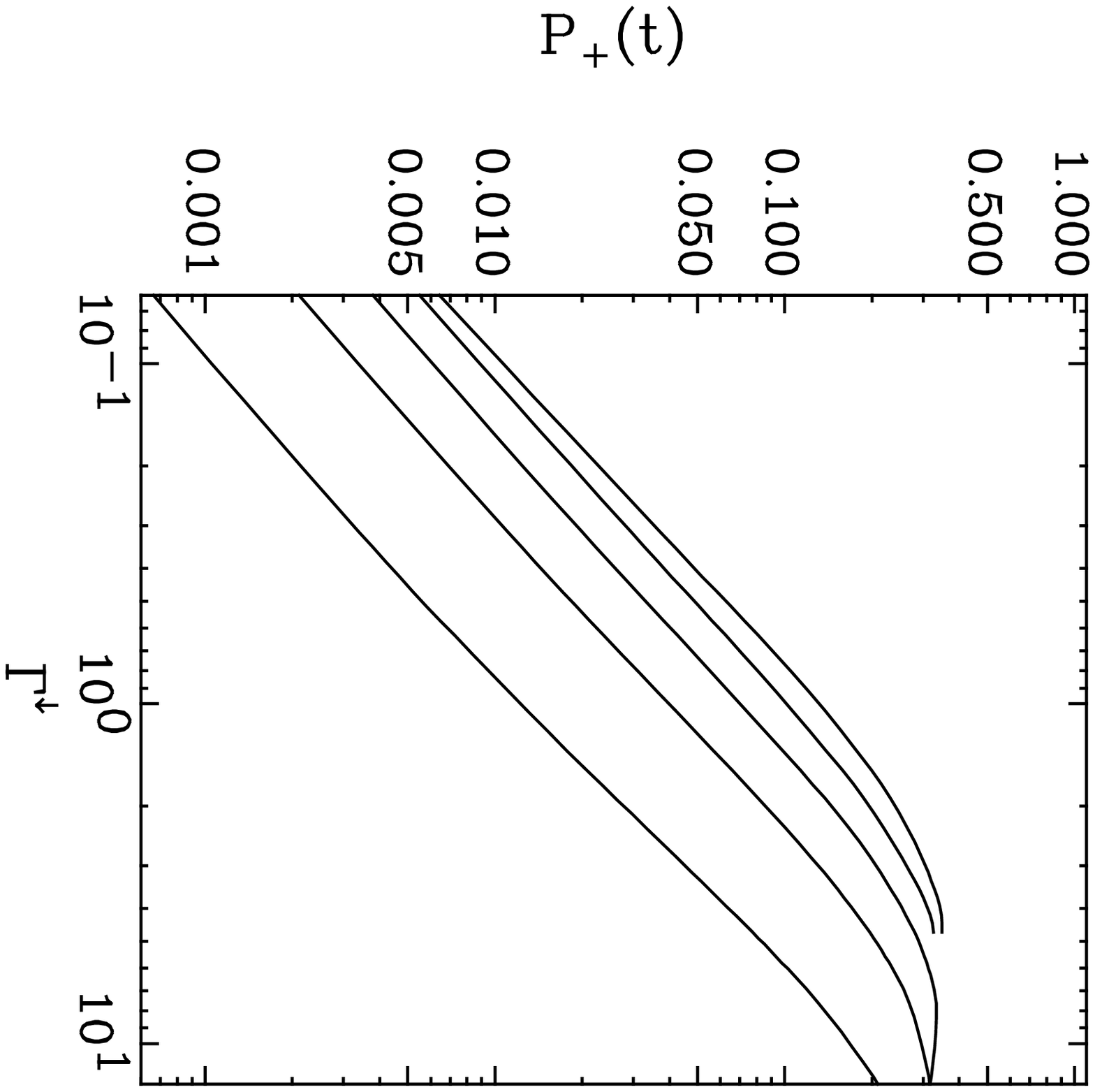 {  The probability to find the particle in the other
well has the approximate power law behaviour $P_+(t)\propto { \Gamma
^\downarrow }^{3/2}$. Different curves correspond to times $t=1$, 2, 3,
4 and 5 respectively in ascending order. } 0.2 3 -.5 1

The role played by the correlation length $X_0$ is however somewhat more
subtle. The rate of tunneling is increasing dramatically when $X_0$ is
decreasing. At the formal level this can be easily understood from the
evolution equation (\ref{evolution}) for the density matrix. In the
$XY$--plane when $|X-Y|<X_0$, the term responsible for dissipation is
negligible. The smaller the correlation length $X_0$ the bigger
is the area where dissipation is directly effective. In the region
$|X-Y|>X_0$ the term $i\Gamma ^\downarrow [G(X,Y)-1]\approx -i\Gamma
^\downarrow $  dominates the time evolution of the density
matrix. Thus, in the long time limit the density matrix tends to become
almost diagonal. This tendency could have been inferred as well from the
time dependence of the cumulants for the momentum distribution, which we
have obtained in the previous two subsections. Since these cumulants
increase with time, that means that the derivatives of the density matrix
$\rho (X,Y,t)$ with respect to $s=X-Y$ become larger and larger for
$X=Y$. 

Another and perhaps a better way to understand the role played by the
correlation length $X_0$ on tunneling is by returning to the initial
picture, see Fig. 1. With decreasing $X_0$ the intrinsic subsystem
undergoes more and more transitions on its way from one well to another,
since the density of (avoided) level crossings per unit length
increases. We remind the reader, however, that trying to understand
the role of the energy exchange between the two subsystems in terms of
isolated jumps at level crossings is not quite correct, as it was discussed at
length in Ref. \cite{bdk_ann}. This also follows from the results
presented in Section II. As we stressed there, had the picture of a
random walk in the energy space been an appropriate one, the
emergence of rather long tails in the energy distribution would have
been rather difficult to explain.

As our numerical results suggest and the structure of the evolution
equation (\ref{evolution}) also seems to confirm, for $t\rightarrow
\infty$ the density matrix has the limit $\rho (X,Y,t) \rightarrow
\delta (X-Y)$, which is likely the only stationary solution of Eq.
(\ref{evolution}). At some point in time however, the adiabatic
approximation we have used to derive the evolution equation
(\ref{evolution}) will cease to be valid, namely when the characteristic
momenta will become of the order of $mX_0\kappa _0/\hbar$. In Section
III we established that for velocities of the order of $X_0\kappa
_0/\hbar$ one has the crossover from the adiabatic to the diabatic
regime, when the ``motional narrowing'' sets in. For higher velocities
dissipation becomes weaker, though not at all negligible. At this
point there is however one essential new element which enters the
evolution of the ``slow'' subsystem, the memory effects. We have
commented on this previously in the text. In the absence of the memory
effects in the dynamics of the ``slow'' subsystem ``thermalization''
between the two subsystems apparently cannot be achieved.

\section{Conclusions} 

We are still quite a distance from solving the problem we have
set out to do in the introduction. Nevertheless we have hopefully succeeded
in clarifying several aspects. Here is the best place to draw the line
and evaluate what we have achieved so far and what remains to be done.

We have developed a formalism, which describes a relatively simple
quantum mechanical system coupled to a ``bath'' of intrinsic
excitations. We solved the dynamical evolution equations for these
systems, without making any uncontrollable approximations or
assumptions. We have analysed two types of problems, each of which being
interesting in its own right and also relevant for understanding various
types of experiments. The first class of systems are the so called
driven systems, when a certain number of externally controlled
parameters are changed. Systems ranging from complex molecules to
quantum dots in variable magnetic or electric external fields can be
studied in this way. One question which we hope we have shed light on,
is the velocity dependence of the diffusion constant. During the years
there have been quite a range of answers to this question, some of them
rather intriguing \cite{diffusion}. 

There are situations, when such parameters become dynamical variables,
as for example in nuclear fission, for which one has to use the second
type of methods described above. Our approach is based on an almost
entirely microscopic description of the intrinsic system, using
parametric random banded matrices. We have established that the dynamics
show some new features. Perhaps the most prominent one is the appearance
of extremely long tails in either energy, momentum or spatial
distributions for the subsystems in interaction and the manifestly
non--Gaussian character of the dissipative dynamics. These features
raise significant doubts concerning the applicability of various
phenomenological transport approaches, such as Fokker--Planck and
Langevin equations \cite{fokker}, to finite many--body quantum systems.

Refinements of details of the present scheme are desirable, for example,
the particular parametrization suggested in Ref. \cite{brink} for the
correlator (\ref{correl}) and a more realistic parametrization for the
energy dependence of the average density of states $\rho (\varepsilon
)$. The introduction of a correlator that is not translationally
invariant, which is well inside the scope of our formalism, allows one
to tackle the problem of the scattering by a localized complex system.

We do not have yet a solution for the two subsystems in interaction
outside the adiabatic limit and consequently, we do not have an
understanding of the dynamical evolution for this very important for
experiments case. The main difficulty here seems to reside in finding a
practical scheme to solve the evolution equations, which are already
known. An extension of the present formalism to open quantum systems (to
include particle emission) is desirable as well. Apparently, from the
formal point of view, the only new element would be the replacement of
the Hermitian Hamiltonian $H_1(X)$ by a non--Hermitian one and that
would not lead to unsurmountable difficulties. A non--Hermitian 
$H_1(X)$ would lead however also to a formalism in which the probability
is not conserved anymore, but alternative approaches can be also
envisioned.

One limitation of our approach is the neglect of the shape dependence of
the average density of states for the intrinsic subsystem $\rho
(\varepsilon ,X)$. This limitation excludes the description of processes
in which mechanical work is exchanged. There is no apparent  technical
or methodological difficulty in taking this additional feature into
account and its implementation should be straightforward.  Dissipative
effects arising from the shape dependence of the average  density of
states alone, i.e. due to the time dependence in  $\rho (\varepsilon ,
X(t))$, are expected to give rise to the  so called one--body type of
friction\cite{onebody} or the wall-formula. A simple example is an ideal
gas in a container, which changes its shape. 

We have not explicitly specified the number of degrees of freedom of the
``slow'' subsystem. The reader might be left with the impression that
our approach is limited to one dimensional ``slow'' motion only.
Actually, that is not the case as it would be clear from a closer
analysis. There is however an element of the dynamics, which would
appear only when the dimensionality is two or higher, namely the
emergence of abelian and/or nonabelian gauge fields \cite{berry,gauge}.
By their definition, gauge fields couple to momenta and thus are
expected to appear in the next to the leading order of an adiabatic
approximation. Naively one would incorrectly conclude that gauge fields
are therefore non--adiabatic in nature. Gauge fields would lead
undoubtly to new and interesting dynamical effects. The generalization
of the present approach to include them seems to raise no technical or
methodological issues. 

There always is the nagging question about a microscopic justification
of the entire parametric random matrix approach and especially about
trying to get an idea about where different key parameters come from,
e.g. $X_0, \kappa _0, \Gamma ^\downarrow$. Obviously, that question lies
outside the scope of this work. We think, however, that one can get a
pretty good idea about the appropriate values for such parameters from
both many--body theoretical models and experiment as well. For example,
one can easily consider an ensemble of many interacting Fermions in
various continuously changing containers and thus extract $X_0$. We do
not see a principial obstacle here. Concerning the applicability of the
very idea of a random matrix, in order to model a many--body system,
that seems to have been answered convincingly
\cite{random,bohigas,molecules,mesoscopic,jack}.
 
A more ambitious goal is to extend the formalism as to attain a more
complete characterization of the two subsystems in interaction. So far
the formalism conserves only the probability. One might want to have a
formalism in which the total energy or other relevant observables, such
as linear and/or angular momentum, is conserved. For example, one can
devise evolution equations for a slightly generalized density matrix,
such as $\rho (X,Y,\varepsilon , t)$, which is obtained by summing over
intrinsic states with a given energy $\varepsilon $ only. This aspect
has been barely touched upon by us \cite{bdk_to_come} and appears to be
a most daunting task. One can write down with relative ease an
impressive number of evolution equations, satisfying such requirements.
However the analysis and the solution of these equations seems at the
present moment quite difficult, to say the least. Another direction,
which seems to be the most difficult technically at the present time, is
to extend the formalism so as to be able to compute fluctuation
characteristics of the dynamics.

We greatly appreciate the DOE support for this work and the computing 
facilities provided by IDRIS and NERSC. The Laboratoire de Physique
Th\'eorique et Hautes Energies is a Laboratoire associ\'e au C.N.R.S.,
URA 0063.


\end{document}